\documentclass[conference,a4paper]{IEEEtran}
\IEEEoverridecommandlockouts

\usepackage{fancyhdr,lipsum}

\fancypagestyle{preprint}{%
   \fancyhf{} 
   
   \fancyhead[L]{PREPRINT}
}%

\makeatletter
\let\ps@IEEEtitlepagestyle\ps@preprint
\makeatother

\usepackage{cite}

\usepackage{amsmath,amssymb,amsfonts}
\usepackage{booktabs}
\usepackage{lipsum}
\usepackage{mathtools}
\usepackage{cuted}
\usepackage{algorithm}
\usepackage{algorithmic}
\usepackage{graphicx}
\usepackage{textcomp}
\usepackage{xcolor}
\usepackage{multirow}
\def\BibTeX{{\rm B\kern-.05em{\sc i\kern-.025em b}\kern-.08em
    T\kern-.1667em\lower.7ex\hbox{E}\kern-.125emX}}
\usepackage{etoolbox}
\include{misc/user_colors}
\makeatletter
\patchcmd{\@makecaption}
  {\scshape}
  {}
  {}
  {}
\makeatother

\usepackage[active=true]{comments}
\newCommentsColor{bharath}{0.0, 0.5, 0.0}
\newCommentsColor{muhammet}{0.0, 0.0, 0.5}
\newCommentsColor{pierre}{0.5, 0.0, 0.0}

\begin{document}

\title{Reducing The Amortization Gap of Entropy Bottleneck In End-to-End Image Compression}


\author{\IEEEauthorblockN{Muhammet Balcilar} \IEEEauthorblockA{\textit{InterDigital, Inc.}\\
Rennes, France \\
muhammet.balcilar@interdigital.com}
\and
\IEEEauthorblockN{Bharath Damodaran}
\IEEEauthorblockA{\textit{InterDigital, Inc.}\\
Rennes, France \\
bharath.damodaran@interdigital.com}
\and
\IEEEauthorblockN{Pierre Hellier}
\IEEEauthorblockA{\textit{InterDigital, Inc.}\\
Rennes, France \\
pierre.hellier@interdigital.com}
}

\maketitle

\begin{abstract}

End-to-end deep trainable models are about to exceed the performance of the traditional handcrafted compression techniques on videos and images. The core idea is to learn a non-linear transformation, modeled as a deep neural network, mapping input image into latent space, jointly with an entropy model of the latent distribution. The decoder is also learned as a deep trainable network, and the reconstructed image measures the distortion. These methods enforce the latent to follow some 
prior distributions. Since these priors are learned by optimization over the entire training set, the performance is optimal \textit{in average}. However, it cannot fit exactly on every single new instance, hence damaging the compression performance by enlarging the bit-stream. In this paper, we propose a simple yet efficient instance-based parameterization method to reduce this amortization gap at a minor cost. The proposed method is applicable to any end-to-end compressing methods, improving the compression bitrate by $1\%$ without any impact on the reconstruction quality. 
\end{abstract}


 \section{Introduction}
\label{sec:intro}


Image and video compression is still a topic of utmost importance, despite decades of hard work. The pandemic, as well as the rise of the metaverse, lead to a huge volume of video transmitted over the network, and limiting the amount of data matters to reduce the energy consumption. Classical methods, based on handcrafted extracted features such as DCT, have prevailed over the last three decades. Since roughly five years, methods based on machine learning have emerged, and their performances now challenge those of traditional techniques. 

These end-to-end deep compression methods are a special case of Variational Autoencoder (VAE) model as in \cite{kingma2013auto}, where the encoder, decoder and entropy model are learned jointly. The output of the encoder is  referred to as the latent variable, and the entropy model prior is trained on the distribution of observed latent variable over the training dataset \cite{theis2017lossy,balle2016end,balle2018variational,minen_joint,minnen2020channel,cheng2020image,xie2021enhanced}. It was shown that minimizing the evidence lower bound (ELBO) of this special VAE is equivalent to minimizing jointly the mean square error (MSE) of the reconstruction image and the entropy of latents w.r.t the priors \cite{balle2018variational}. 

At test time, these codecs encode the quantized latent variable losslessly  \textit{w.r.t} the trained priors into a bit-stream. It is typically performed by any entropy coder such as range or arithmetic coding \cite{1056282}. The prior distribution being known by the decoder (or reconstructed by side information \cite{balle2018variational,minen_joint,minnen2020channel, cheng2020image,xie2021enhanced}), the latent can be reconstructed from the transmitted bitstream, and consequently used to reconstruct the original data. At the moment, various priors have been proposed so far: fully-factorized \cite{balle2016end}, zero-mean gaussian \cite{balle2018variational}, gaussian \cite{minen_joint,minnen2020channel} or mixture of gaussian \cite{cheng2020image}, where some predict priors using an autoregressive schema \cite{minen_joint,minnen2020channel,cheng2020image,xie2021enhanced}. 

All prior models are learned by amortizing its parameters over the entire training set. Consequently, the learned prior is optimal \textit{in average}, but sub-optimal for a given specific data instance. This is due to the fact that the observed latent distribution differs from the learned prior distribution. This problem is referred to as the amortization gap \cite{cremer2018inference}. Methods proposed to solve this problem are two folds: firstly, enforce that the latent of the given instance obeys the priors \cite{Aytekin_2018_CVPR_Workshops,LuCZCOXG20,Campos_2019_CVPR_Workshops,NEURIPS2020_066f182b,Guo_2020_CVPR_Workshops}; secondly, modify the priors to better fit the given instance's latents \cite{9088301,Lam_Yat_Hong,9287069,van2021overfitting}. The first class of methods does not need any update on the receiver side, but has limited gain. The second class of methods can update the encoder/decoder in addition to entropy model as well resulting more gain, at the additional cost of transmitting these updates to the receiver. However, all methods require post-training to overfit on a given data instance, which increases the encoding time significantly.

In this paper, we propose two main contributions: first, we define the amortization gap of entropy model in compressing perspective and report the amortization gap of some recent neural image compression model over benchmark datasets. Second, we propose simple yet efficient methods for factorized and hyperprior entropy models to adjust the priors to fit on any new instance to be compressed. Our solution does not need post-training and does not add computational complexity. We show that a gain of at least $1\%$ can be expected from sota end-to-end compression method without any impact on the reconstruction quality.

\section{Neural Image Compression}
\label{sec:endtoend}

In this section, we introduce the mathematical notations for the end-to-end variational autoencoders and the models of the entropy model. The encoder $\mathbf{y}=g_a(\mathbf{x};\mathbf{\phi})$, where $\mathbf{\phi}$ are the parameters of the corresponding neural network, transforms each input $\mathbf{x} \in \mathbf{R}^{n\times n \times 3}$ into a lower dimensional latent $\mathbf{y} \in \mathbf{R}^{m\times m \times o}$ and quantize it to obtain $\mathbf{\hat{y}}=\mathbf{Q}(\mathbf{y})$.

Recent methods have proposed the two following possible solutions:
\begin{itemize}
    \item The fully factorized model \cite{balle2016end}, where the the quantized latents are compressed losslessly by the entropy coder using factorized entropy model $p_{f}(\mathbf{\hat{y}}|\Psi)$.
    \item The hyperprior model, now used in most settings \cite{balle2018variational,cheng2020image}, where a side information is extracted so as to remove spatial structure from the latent information so that the model generalizes better. The side information $\mathbf{z}=h_a(\mathbf{y} ;\Phi)$ where $\mathbf{z} \in \mathbf{R}^{k\times k \times f}$ (and its quantization $\mathbf{\hat{z}}=\mathbf{Q}(\mathbf{z})$) are also learned. In that case, $\mathbf{\hat{y}}$ is encoded with the hyperprior entropy model $p_{h}(\mathbf{\hat{y}}|\mathbf{\hat{z}};\Theta)$, and $\mathbf{\hat{z}}$ is encoded with factorized entropy model $p_{f}(\mathbf{\hat{z}}|\Psi)$.  
\end{itemize}

The decoder $\mathbf{\hat{x}}=g_s(\mathbf{\hat{y}};\mathbf{\theta})$ reconstructs the image $\mathbf{\hat{x}}$ from the transmitted quantized latent variables, or reconstructed latent in the case of the hyperprior model. In the general case, the parameters $\mathbf{\phi}, \mathbf{\theta}, \Psi, \Theta $ of $g_a,g_s,p_{f}, p_{h} $ are obtained by minimizing the following rate-distortion loss.
\begin{equation}
\small
   \label{eq:balle2}
   \mathcal{L}=\mathop{\mathbb{E}}_{\substack{\mathbf{x}\sim p_x \\ \epsilon \sim U}}\left[-log(p_{h}(\mathbf{\hat{y}}|\mathbf{\hat{z}},\Theta)) -log(p_{f}(\mathbf{\hat{z}}|\Psi)) + \lambda d(\mathbf{x},\mathbf{\hat{x}})\right],
\end{equation}
where $d(.,.)$ is any distortion loss such as MSE, $\lambda$ is the trade-off parameter to control compression ratio and quality, $\mathbf{Q}(.)$ is continuous relaxation at train time as $\mathbf{Q}(x)=x+\epsilon$, $\epsilon \sim U(-0.5,0.5)$. 

At test time, the quantized latent variables are compressed losslessly by the entropy coder as follows:
\begin{itemize}
    \item For the fully factorized model \cite{balle2016end}, each $k \times k$ slice of side latent has a trainable cumulative distribution function (cdf) in entropy model shown by $\bar{p}_{\Psi}^{(c)}(.), 
~c=1\dots f$, 
and probability mass function (pmf) for a given value of $x$ is derived 
as $\hat{p}_{\Psi}^{(c)}(x)=\bar{p}_{\Psi}^{(c)}(x+0.5)-\bar{p}_{\Psi}^{(c)}(x-0.5)$.
Thus, the entropy model applies as follows; 
\begin{equation}
   \label{eq:factorent}
   p_{f}(\mathbf{\hat{z}}| \Psi)=\prod_{c=1}^{f} \prod_{i,j=1}^{k,k} \hat{p}_{\Psi}^{(c)}({\mathbf{\hat{z}}_{i,j,c}})
\end{equation}
\item In the case of the hyperprior model, the entropy model of the latent $\mathbf{\hat{y}}$ is conditioned with the side information $\mathbf{z}=h_a(\mathbf{y} ;\Phi)$. Thus, $\mathbf{\hat{y}}$ is encoded with the hyperprior entropy model $p_{h}(\mathbf{\hat{y}}|\mathbf{\hat{z}};\Theta)$, and $\mathbf{\hat{z}}$ is encoded with factorized entropy model $p_{f}(\mathbf{\hat{z}}|\Psi)$.
\end{itemize}

Let us describe in more details the special case of the hyperprior model. Each latent point is modeled as $1d$ Gaussian distribution and its pmf is $\hat{N}(x;\mu,\sigma)=\bar{N}(x+0.5;\mu,\sigma)-\bar{N}(x-0.5;\mu,\sigma)$ while $\bar{N}(.;\mu,\sigma)$ is the cdf of $1d$ Gaussian distribution. The hyperprior entropy model is written $p_{h}(\mathbf{\hat{y}}|\mathbf{\hat{z}},\Theta) = \prod_i \hat{N}(\mathbf{\hat{y}}_i;\mu_i,\sigma_i)$at train time where $\mu,\sigma=h_s(\mathbf{\hat{z}};\Theta)$ as in \cite{balle2018variational} 
or  
$\mu_i,\sigma_i=h_s(\mathbf{\hat{z},\mathbf{\hat{y}}_{<i}};\Theta)$ in autoregressive prediction in \cite{minen_joint,cheng2020image,xie2021enhanced}. $h_s$ is a trainable model implemented as a neural network with parameter $\Theta$.
 However, this implementation is not effective at test time due to the necessity of recalculating the pmf table 
at receiver side for each latent points $i$.
Thus, it is common practice to use $s$ number of predefined integer resolution pmf tables 
under zero means but different scale parameters (logarithmic distributed scale values between $\sigma_{min}$ to $\sigma_{max}$)
\cite{balle2018variational,minen_joint,minnen2020channel, cheng2020image,xie2021enhanced}. As long as $\mathbf{\Tilde{y}}_i=Q(\mathbf{y}_i-\mu_i)$, $\mathbf{\hat{y}}_i=\mathbf{\Tilde{y}}_i+\mu_i$, $\sigma_c$ is $c$-th predefined scale and  $\mathcal{N}(\sigma_c)$ is a set of latent indices whose winning scale is $\sigma_c$, the hyperprior entropy model is implemented as follows at test time:
\begin{equation}
   \label{eq:hyperentapp}
   p_{h}(\mathbf{\hat{y}}|\mathbf{\hat{z},\Theta}) = \prod_{c=1}^{s} \prod_{i \in \mathcal{N}(\sigma_c)} \hat{N}(\mathbf{\Tilde{y}}_i;0,\sigma_c)
\end{equation}

\section{Proposed Method}
\label{sec:proposal}

In this section, we define the amortization gap of the entropy models and propose solutions to reduce it.

\subsection{Amortization Gap of the Entropy Model}
\label{ssec:gap}
At training time, parameters $\mathbf{\phi,\theta},\Phi,\Theta$ and $\Psi$ are estimated by optimizing  $\mathcal{L}$ over the data distribution $\mathbf{x} \sim p_x$. 
The parameters may be optimal \textit{in average} for entire dataset, but not any specific instance $\mathbf{x}$, which is known as an amortization gap \cite{cremer2018inference}. Actually, the amortization gap in compression schema may occur for each trainable blocks in the model. However, we are here only interested in the gap for the entropy models.

The amortization gap of the entropy model is the difference between the optimal entropy model and the learned entropy model (see Fig.~\ref{fig:gap}). It quantifies the expected gain in bit length, if the entropy models' pmfs are optimal on every input instance.
This gap can be calculated for the factorized entropy model as follows:
\begin{equation}
   \label{eq:gap1}
   \mathcal{G}_f=-log(p_{f}(\mathbf{\hat{z}}|\Psi))   +\log(p^{*}_{f}(\mathbf{\hat{z}}))
\end{equation}
$p^{*}_{f}(\mathbf{\hat{z}})$ refers to the optimal entropy model within the same entropy family $\mathcal{P}_f$. 
For the hyperprior entropy model, it reads: 
\begin{equation}
   \label{eq:gap}
   \mathcal{G}_h=-log(p_{h}(\mathbf{\hat{y}}|\mathbf{\hat{z}},\Theta)) +log(p^{*}_{h}(\mathbf{\hat{y}})) 
\end{equation}
$p^{*}_{h}(\mathbf{\hat{y}})$ refers to the optimal entropy model within the same entropy family $\mathcal{P}_h$. 
Eqn.~\ref{eq:gap} and \ref{eq:gap1} shows that this gap can be bridged by using the optimal entropy model on each instance. To find such an instance-specific optimal entropy model, one does not need to optimize the log-likelihood, since the normalized histogram is the optimal pmf \cite{mezard2009information}. 

Thus, replacing learned pmf tables with histogram of actual latents for each instance removes the amortization gap of the entropy model. However, it introduces the additional cost of transmitting these histograms for each instance, enlarging the bit stream which is not practically feasible.

\subsection{Explicit Parameterization}
\label{ssec:prop}

\begin{figure}[htb]
\includegraphics[width=0.97\linewidth]{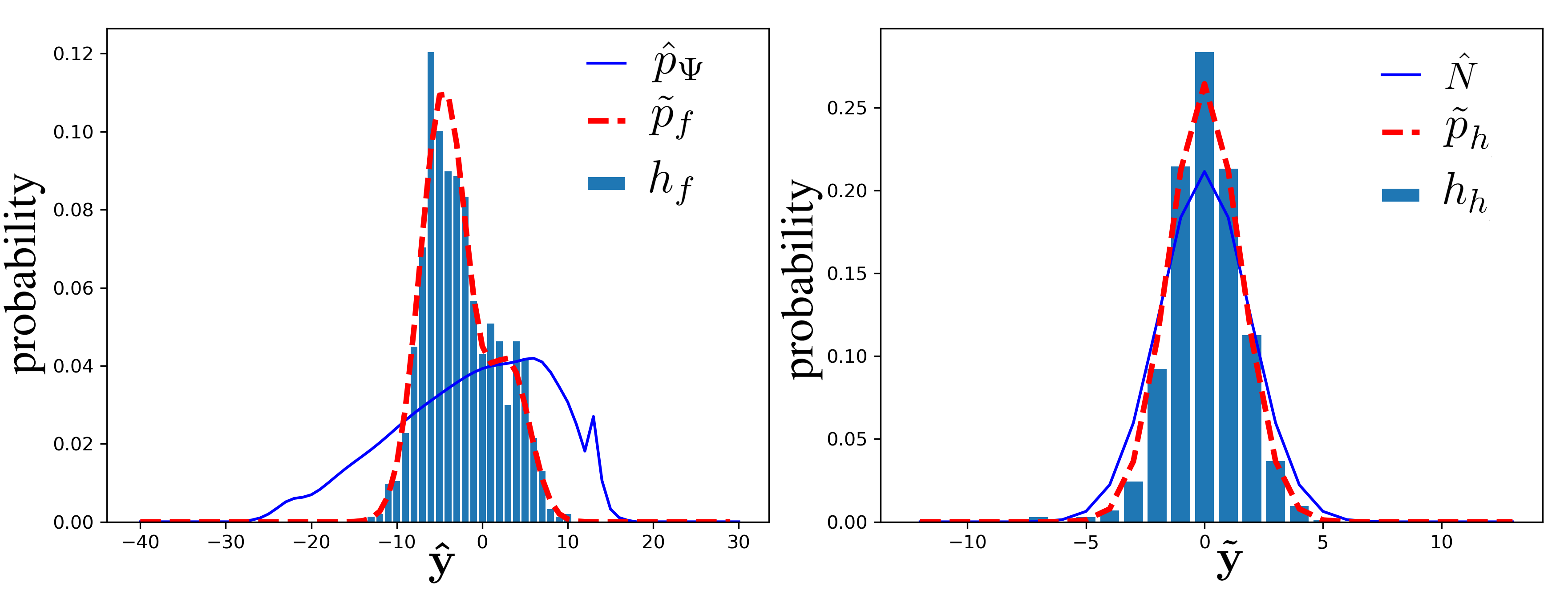}
\vskip -0.15in
\caption{Learned pmfs ($\hat{p}_{\Psi},\hat{N}$), reparameterized pmfs ($\Tilde{p}_f,\Tilde{p}_h$) and normalized frequencies ($h_f,h_h$) for a certain image's selected latent under factorized (left) and hyperprior (right) entropy models. Our reparameterization fits better on the normalized frequencies, leading to improved compression.}
\label{fig:gap}
\end{figure}

Here, we propose an efficient solution to bridge the amortization gap of the entropy model at a negligible additional transmission cost by an explicit parameterization. More specifically, when an input image has to be encoded, we parameterize the distribution of the latents $\Tilde{p}_{f|h}(.,\beta)$ for factorized and hyperprior entropy models to closely approximate the optimal pmf. Our approximation is illustrated in Fig. \ref{fig:gap}. This low-level approximation can be hopefully transmitted at a negligible extra cost (signaling cost of $\beta$), and aims at improving the encoding using $\Tilde{p}_{f|h}(.,\beta)$.  

We propose two variants for approximating $\Tilde{p}_{f|h}(.,\beta)$. The first one is generic, uses Gaussian mixture model and is described in section \ref{sec:GMM}. It is dedicated to the factorized entropy model, since this modeling is flexible. The second one is a simplified version, where the central bin is spread on neighbouring bins, and is described in section \ref{sec:diffbin}. This approach is dedicated to the hyperprior entropy model, since the discrepancy between learned and actual pmf is smaller.

\subsubsection{Truncated Gaussian Mixture on Discrete Support}
\label{sec:GMM}

Since the factorized entropy model is a non-parametric distribution model \cite{balle2016end,balle2018variational}, the pmf is flexible enough to have any shape. One of the most successful parametric distribution model is
Gaussian Mixture Model (GMM) which can approximate any smooth density function \cite{goodfellow2016deep} with a cost of three parameters per component.
Thus, we propose to use GMM as a tool to re-model the latent distribution of the factorized entropy model.

In our case, the function to be approximated is defined on integer center and support domain of GMM is truncated such that $[x_{min} \dots x_{max}]$. Thus, we can write our mixture model's pmf for factorized entropy model $\Tilde{p}_f(x;\beta^{(c)}))$ as:
\begin{equation}
   \label{eq:truncatedgmm}
   \Tilde{p}_{f}(x;\beta^{(c)}) =\frac{\sum_{k=1}^{K}\pi_k N(x;\mu_k,\sigma_k)}{\sum_{z=x_{min}}^{x_{max}} \sum_{k=1}^{K}\pi_k N(z;\mu_k,\sigma_k)}
\end{equation}
\noindent Here $\beta^{(c)} =\{\pi_k,\mu_k,\sigma_k\}_{k=1 \dots K}$ denotes the set of parameters to be inferred in \eqref{eq:truncatedgmm} for $c$-th latent band.
Since there is no closed form solution of $\beta^{(c)*}$ that maximize \eqref{eq:truncatedgmm}, optimization is used for estimation. In practice, given the small number of parameters, convergence is fast. Fig \ref{fig:gap} (left) displays the result of pmf approximation using a mixture of two Gaussians ($\scriptstyle K=2$).
In practice, the tuning of parameter $K$ leads to a trade-off between approximation accuracy and transmission cost.


For the hyperprior entropy model, since the latent distribution has zero-mean and is uni-modal, we relax eqn.~\ref{eq:truncatedgmm} by setting $K=1$ and $\mu=0$, also called as zero-mean truncated Gaussian approximation. This is illustrated in Fig.~\ref{fig:gap}(right) where $\Tilde{p}_{h}(x;\beta^{(c)})$ obtained by equation \eqref{eq:truncatedgmm} under $K=1$ and $\mu=0$.


\subsubsection{Difference of the Center Bins Probability}
\label{sec:diffbin}
We propose in this section a simpler alternative for the hyperprior entropy, since the latter is parametric and by construction has a zero-mean Gaussian shape. Hence, in that special case, we propose an alternative at minimal transmission cost. We use the heuristic of computing the error of center bins  probability and spread this error to the other bins proportionally seems arguable strong closed form alternative. In this approach, reparameterized $c$-th pmf of hyperprior entropy model can be written in \eqref{eq:diffcenter} where the parameter $\beta$ is the error of the center bins probability between learned one and the actual one in the normalized histogram such as   $\beta^{(c)}=\hat{N}(0;0,\sigma_c)-h_h^{(c)}(0)$.

\begin{equation}
\small
   \label{eq:diffcenter}
  \Tilde{p}_h(x;\beta^{(c)}) =
  \begin{cases} 
      \hat{N}(x;0,\sigma_c)-\beta^{(c)} & \mbox{if }x = 0 \\
      \hat{N}(x;0,\sigma_c)  (1+\frac{\beta^{(c)}}{1-\hat{N}(0;0,\sigma_c)}) & \mbox{if }x \neq 0
   \end{cases}
\end{equation}

\subsubsection{Quantization of the Parameters}
\label{sec:quantn}
According to selected re-parameterization method, 4 different kind of parameters ($\mu, \sigma, \pi$ of Gaussian and differences of center bin probability) should be explicitly encoded into the bitstream as an extra parameters. In order to do that, we apply quantization on predefined number of bins (8-bit by default). Since the range of these parameters are different, we prepared shared quantization tables for each parameters. We discretize $\sigma$ in range of $[0.002,20]$ with logarithmic bin width, $\pi$ in range of $[0,1]$, $\mu$ in range of $[x_{min},x_{max}]$ (minimum symbol to maximum symbol in the cdf table) and differences of center bin probability in range of $[-0.03,0.03]$ with uniform bin width using 256 quantization centers  for 8-bit quantization. 
\section{Experimental Results}
\label{sec:result}

We used CompressAI  
library \cite{compressai} to test our contributions on already implemented $6$ SOTA deep compression methods, as well as on a very recent method \cite{xie2021enhanced}. We used two datasets to evaluate our method: Kodak test set\cite{eastman_kodak_kodak_nodate} and Clic-2021 Challenge's Professional test set \cite{CLIC}, consisting of 24 and 60 images respectively. 

\noindent \textbf{Analysis of amortization gap:} We measured the amortization gaps of pre-trained SOTA methods and our gains on Kodak test set \cite{eastman_kodak_kodak_nodate} on the lowest bpp. Results are given in Table~\ref{res1}.
Regardless of the baseline method, the factorized entropy model's amortization gap is quite large (7.6-11.8\%), compared to the hyperpriors one (1.9-4.5\%). This observation is also predictable by Fig~\ref{fig:gap} where it clearly shows that mismatch between $h_f$ and $\hat{p}_{\Psi}$ is much more bigger than $h_h$ and $\hat{N}$. 
This can be explained easily: the hyperprior entropy model uses instance specific information, and fully factorized model does not. The hyper-prior methods encode very small amount of the data (0.6-5.9\%) with less effective entropy model (factorized entropy), but vast majority of them (94.1-99.4\%) is encoded by effective entropy model (hyperprior entropy). Thus, in average their amortization gap (1.9-4.7\%) is smaller compared to the fully-factorized method (9.5\%).
From the different version of same method, it can be seen that when the amount of side information decreases, correspondingly the hyperprior gap increases. For instance, the hyperprior gap is 3.4\% where there is 5.9\% side information 
but increases to 4.5\% when there is 3.5\% side information. 

\textbf{Reduction of Amortization Gap: }Our proposed methods managed to save significant amount of bits on all studied methods in Table~\ref{res1}. The magnitude of saving is higher with the factorized entropy model compared to hyper-prior model. We can reduce the factorized entropy gap from 9.5\% to 2.71\%, with the gain of 6.79\% when all information is encoded by this entropy model in \textbf{bmshj2018-factorized} model. However, when the amount of information to be encoded is less with this entropy model, our gain also reduces. For instance, when only 2.3\% of the information is encoded by this entropy model in \textbf{cheng2020-attn}, our gain is 2.77\%. The fixed extra parameter cost in our proposal makes the efficiency of our method proportionally less when the amount of information to be encoded decreases. The hyperprior entropy encodes large percentage of the information, and our gain ranges from 1-1.8\%, which almost fills in average 45\% gap exists in hyperprior entropy over different methods.

\begin{table*}[tb]
\caption{Ratio of the encoded information, the amortization gap of the entropy models and our gain for each entropy model relative to the original bit-length for the methods trained lowest bpp objective. The gap and the gain is much more higher in factorized entropy. Our solution reduces the gap significantly both factorized and hyperprior entropy model. In total, proposed method saves more than 1\% of file size from sota model. Results are averaged over Kodak Test set.
}
\label{res1}
\begin{center}
\begin{scriptsize}
\begin{sc}
\begin{tabular}{lccccccccc}
\toprule
Model & \multicolumn{3}{c}{Factorized Entropy} & \multicolumn{3}{c}{Hyperprior Entropy}  & \multicolumn{2}{c}{Total} \\
 & Ratio &  Gap & Gain & Ratio & Gap & Gain & Gap & Gain\\
 & (in \%) & (in \%) & (in \%) & (in \%) & (in \%) & (in \%) & (in \%) &(in \%)  \\
\midrule
bmshj2018-factorized \cite{balle2016end}    &100 & 9.5 & 6.79 &- &- &- & 9.5 & 6.79 \\
bmshj2018-hyperprior \cite{balle2018variational}  &3.5 &   10.0 & 4.65 & 96.5&  4.5  & 1.84 & 4.7 & 1.98\\
mbt2018-mean  \cite{minen_joint}    & 5.9 &  11.4& 4.27 & 94.1 & 3.4 &1.24 & 3.8 & 1.43\\
mbt2018 \cite{minen_joint}    &  2.3 & 9.9 & 4.09 & 97.7 & 2.5  & 0.99 & 2.7 & 1.06\\
cheng2020-anchor \cite{cheng2020image}   & 1.2  &   11.8  & 3.68 & 98.8 & 2.3  & 1.06 & 2.5 & 1.09\\
cheng2020-attn \cite{cheng2020image}      & 2.3&  9.5 &  2.77 &  97.7 &1.9  & 0.88 & 2.1 & 0.92\\
InvCompress  \cite{xie2021enhanced}      & 0.6 &  7.6 & 2.18  & 99.4 & 3.0 & 1.32 & 3.0 & 1.33\\
\bottomrule
\end{tabular}
\end{sc}
\end{scriptsize}
\end{center}
\end{table*}

\begin{figure}[tb]
\includegraphics[width=1.0\linewidth]{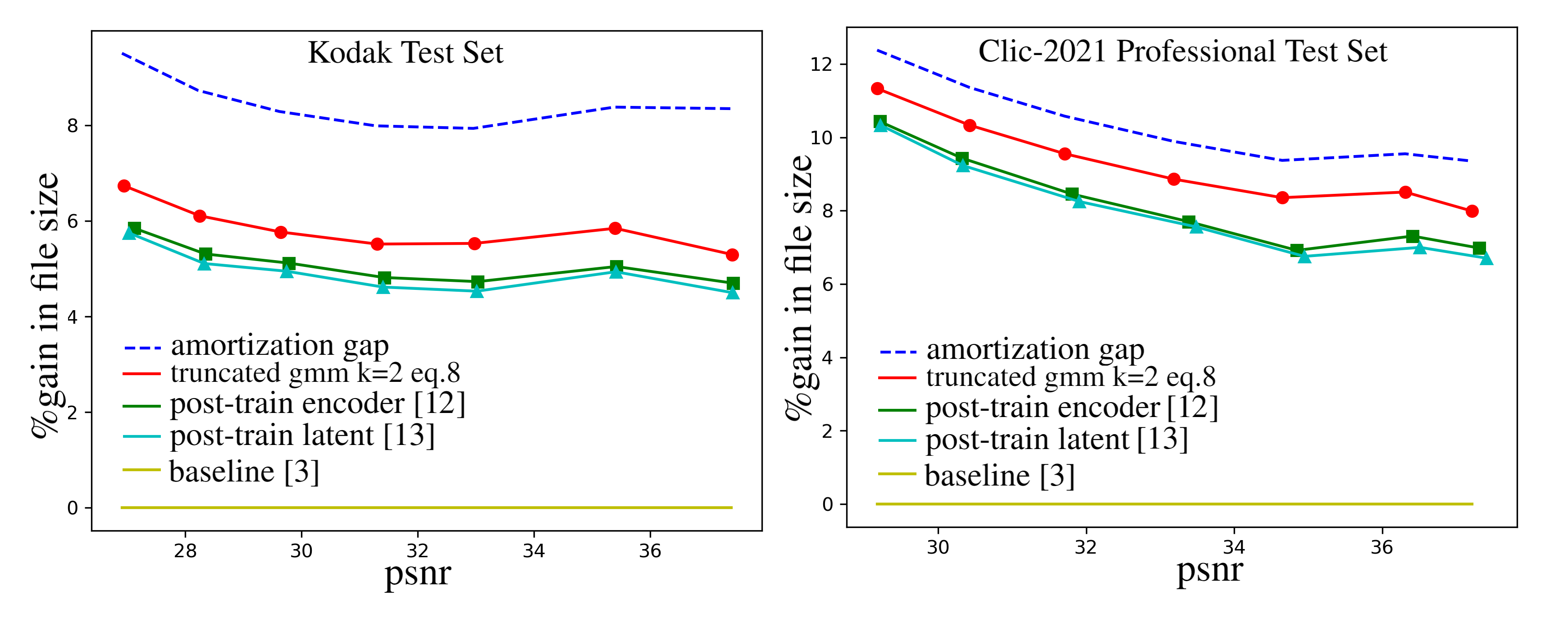}
\vskip -0.10in
\caption{Experimental results on Kodak and Clic Test Set for \textbf{bhshj2018-factorized} model in \cite{balle2016end} trained on 7 different psnr objectives. Proposed method based on truncated GMM saves 6.8\% original bit length for Kodak, 11.5\% for Clic test test on lower psnr and outperformed post-train based high computational demanding alternatives.}
\label{fig:res1}
\vskip -0.10in
\end{figure}

\begin{figure}[tb]
\includegraphics[width=1.0\linewidth]{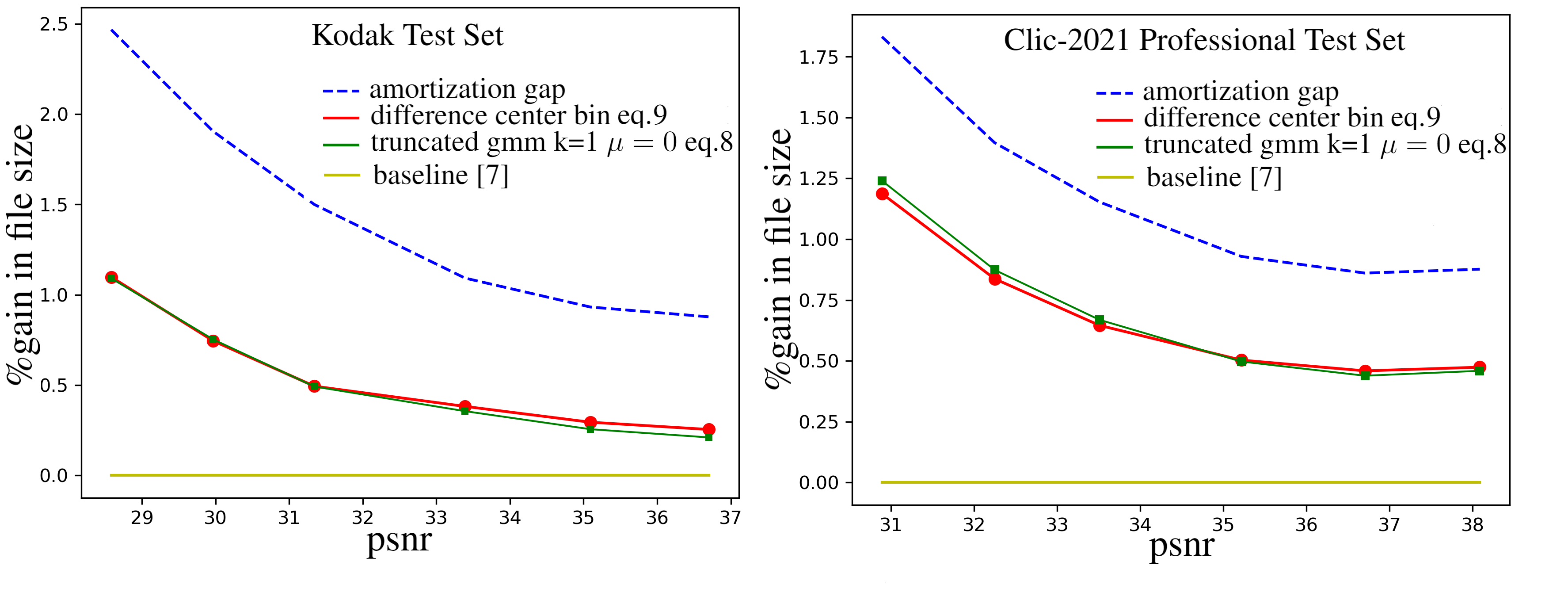}
\vskip -0.10in
\caption{Experimental results on Kodak and Clic Test Set for \textbf{cheng2020-anchor} model in \cite{cheng2020image} trained on 6 different psnr objectives. The gap and our gain decrease on higher psnr target. Proposed two methods 
save more than 1\% of file size for both test sets at lower psnr. Our simple center bin difference solution gives comparable results even better at higher psnr.}
\label{fig:res2}
\vskip -0.10in
\end{figure}

In order to measure the performance of proposed method on different PSNR targets, we plugged our method on the pre-trained model in \cite{balle2016end} for the factorized entropy and the best neural compressing model \textbf{cheng2020-anchor} \cite{cheng2020image} provided in \cite{compressai} for the hyperprior entropy.The performances are measured with Kodak and Clic-2021 Challenge's Professional test set. According to results in Fig~\ref{fig:res1}, the amortization gap of factorized entropy varies from 8.5\% to 9.5\% in Kodak dataset where the proposed method gains from 5.3\% to 6.8\% in file size. In Click-2021 dataset, the gap (9.5\%-12.5\%) and our gain (8\%-11.5\%) are even bigger. Fig.~\ref{fig:res2} reports the results of the hyperprior entropy, our method saves more than 1\% of original file size in lower bit-rate and save around 0.5\% in highest bit-rate. The simplest approach that parameterizes the new probability by the difference between center bin's probability in Eq.\ref{eq:diffcenter} gives competitive result even better in higher psnr with zero-mean Gaussian parameterization.

\noindent \textbf{Comparison with competitors:} To compare our method with the existing method, we also implemented two instance based post-training methods: post-train encoder (trains the encoder for given test image) \cite{LuCZCOXG20}, post-train latent (learns more effective instance's latent directly without training encoder)\cite{Campos_2019_CVPR_Workshops}. According to our test, we have found that \cite{Campos_2019_CVPR_Workshops} is faster (but still needs significant time to train) than \cite{LuCZCOXG20} but with less performance as can be seen in Fig~\ref{fig:res1}. Our proposal reaches better results and outperforms significantly compared to these two approaches even without giving any significant computational complexity. 


\noindent \textbf{Computational Complexity:} To test the computational complexity, we use python library of PAPI \cite{papi} and count floating point operations per pixel in flops/pxl both encoding and decoding time. As the results are shown in Table \ref{res2}, our proposal gives negligible extra complexity both encoding and decoding time. Even in the fully-factorized model which uses relatively costly GMM model fitting with $K=2$, our extra complexity demanding is less than \%0.3 in encoding time. Since the hyperprior model uses zero-mean gaussian fitting and has good initial guess with learned standard deviation, extra encoding operation even goes below \%0.05. Since in the decoding time, extra process is to re-arrange pmf table with encoded extra parameters, there is almost no differences in decoding complexity. Note that, our alternatives which applies finetuning in encoding time as in \cite{Campos_2019_CVPR_Workshops,LuCZCOXG20} needs enormous extra computational complexity in encoding. Their extra computational demand is depend on the maximum iteration of finetuning, where each iteration is not less than a single decoding (one forward pass) complexity in any model (most of the time 3-4 times longer than one forward-pass because of necessity of back-propagation of gradients). In our test, we did not get meaningful result from \cite{Campos_2019_CVPR_Workshops,LuCZCOXG20} without applying 1000 iteration which has 3000-4000 times bigger computational demand. 

\noindent \textbf{Implementation Details:} In order to reproducibility, we implement the proposals as a collection of classes of python end-to-end image compression library CompressAI \cite{compressai} and integrate to it as a part of this well-known library. Using our classes, the one can easily apply proposal gap reduction to any model and decrease the file size without any impact on the reconstruction quality. In our models, the re-parameterization method which can be either zero-mean gaussian, GMM with K=1,2,3 or difference of center bin approaches is hyperparameter for both factorized and hyperprior entropy models. In practice, a table-wise selection mechanism is used to determine if the learned pmf should be replaced by the approximated one. This results necessity of signaling receiver by 1-bit if the cdf table is replaced or not. Instead of testing all pmf tables in both entropy models, we define another hyperparameter which shows how many cdf tables are targeded. The number of bits to encode single extra parameter explicitly can also be seen as hyperparameter which is common for both entropy models. We use GMM for $K=2$ and targetted top 64 pmf tables in \textbf{bmshj2018-factorized} model. For \textbf{cheng2020-anchor}, we use GMM for K=1 and targeted 32 pmf for factorized entropy (side information), while zero mean gaussian and targetted 32 pmf for hyperprior entropy (main information). We always use 8-bit quantization on the extra parameters.          

\begin{table}[tb]
\caption{Necessary number of floating point operations per pixel (flops/pix) in encoding and decoding for studied two neural models. Our results are obtained over Kodak test set with our most efficient settings in terms of compression performance.
}
\label{res2}
\begin{center}
\begin{scriptsize}
\begin{sc}
\begin{tabular}{lccccc}
\toprule
Model & \multicolumn{2}{c}{Encoding} & \multicolumn{2}{c}{Decoding}  \\
 & Baseline & Ours &  Baseline & Ours   \\  
\midrule
bmshj2018-factorized \cite{balle2016end}    & 83,797 & 84,024 & 85,400 &85,412\\
cheng2020-anchor \cite{cheng2020image}   & 348,142 & 348,298 & 519,626 & 519,639\\

\bottomrule
\end{tabular}
\end{sc}
\end{scriptsize}
\end{center}
\end{table}

\section{Conclusion}
\label{sec:conclusion}
We have proposed here to improve the efficiency of entropy models in deep neural image compression algorithm. First, we have defined the amortization gap for entropy models, and we measured experimentally the gap for sota methods. Then, we have proposed an effective and computationally-friendly method to fill the gap, which differs from previously published methods. We have shown experimentally that a gain above $1\%$ was obtained, on different dataset and on $7$ various SOTA compression methods. Our method can actually be applied to any end-to-end deep compression technique, without any impact on the reconstruction quality.
One known limitation of deep methods is reconstruction failure, due to architecture discrepancy (the architecture used at encoding and decoding differ). This is explained by the need of recomputing the pmf table at decoding, and small errors occur because of floating-value approximations \cite{balle2018integer}. In this regards, our method based on center-bin difference will be robust since it needs only integer divide operation. 
Future work may explore the link between explicit parameterization of the entropy model, and fine-tuning of the encoder. We may expect to increase the reconstruction quality, at constant bit-rate.

\bibliographystyle{IEEEtran}
\bibliography{string}


\end{document}